# High fidelity optogenetic control of individual prefrontal cortical pyramidal neurons *in vivo*

Shinya Nakamura, Michael V Baratta, Matthew B Pomrenze, Samuel D Dolzani, Donald C Cooper*

Department of Psychology and Neuroscience, Institute for Behavioral Genetics, University of Colorado, Boulder, CO 80303 USA

*Corresponding author: Donald C Cooper, email: d.cooper@colorado.edu

**Abstract** Precise spatial and temporal manipulation of neural activity in specific genetically defined cell populations is now possible with the advent of optogenetics. The emerging field of optogenetics consists of a set of naturally-occurring and engineered light-sensitive membrane proteins that are able to activate (e.g., channelrhodopsin-2, ChR2) or silence (e.g., halorhodopsin, NpHR) neural activity. Here we demonstrate the technique and the feasibility of using novel adeno-associated viral (AAV) tools to activate (AAV-CaMKIIα-ChR2-eYFP) or silence (AAV-CaMKIIα-eNpHR3.0-eYFP) neural activity of rat prefrontal cortical prelimbic (PL) pyramidal neurons *in vivo*. *In vivo* single unit extracellular recording of ChR2-transduced pyramidal neurons showed that delivery of brief (10 ms) blue (473 nm) light-pulse trains up to 20 Hz via a custom fiber optic-coupled recording electrode (optrode) induced spiking with high fidelity at 20 Hz for the duration of recording (up to two hours in some cases). To silence spontaneously active neurons we transduced them with the NpHR construct and administered continuous green (532 nm) light to completely inhibit action potential activity for up to 10 seconds with 100% fidelity in most cases. These versatile photosensitive tools combined with optrode recording methods provide experimental control over activity of genetically defined neurons and can be used to investigate the functional relationship between neural activity and complex cognitive behavior.

## Introduction

A method for selective and rapid reversible manipulation of neuronal activity is important for parsing out the relationship between prefrontal cortical (PFC) neuronal activity and cognition. Recent optogenetic technologies, which allow neurons to respond to specific wavelengths of light with action potential output, are now providing a significant advance in our ability to control the activity of select cell populations and neural circuits in behaving animals[1]. These tools allow for bidirectional control over the neuronal activity[2]. To date, the majority of optogenetic experiments have used transgenic, virally-mediated, or a combination of the two approaches in mice[3,4,5]. Here we use virally-mediated gene delivery of the light-responsive proteins ChR2[6] or halorhodopsin[7] in Sprague-Dawley rat PFC PL pyramidal neurons. Expression of ChR2 and NpHR enables neurons to be depolarized and silenced by pulses of blue and green light, respectively. We tested how well PL pyramidal neurons expressing ChR2 followed pulses (10 ms) of blue light delivered at 20 Hz (fidelity). Given the potential importance of long duration light delivery for a variety of protocols we determined if stable responses could be elicited from individual neurons for at least 2 hrs. Lastly, we tested green light-induced silencing of spontaneously active NpHR expressing PL neurons to determine if network-driven action potential activity could be silenced continuously for 10 seconds, thus establishing the feasibility of long duration silencing for behavioral testing.

## Materials and methods

*Subjects.* Male Sprague-Dawley rats (6–8 weeks) were housed in pairs on a 12-h light/dark cycle. Rats were allowed to acclimate to colony conditions for 7–10 days prior to surgery. All animal procedures were approved by the Institutional Animal Care and Use Committee of University of Colorado at Boulder.

*Virus injection.* An AAV vector carrying the opsin gene encoding the light-gated nonselective cation channel ChR2 or the light-driven third generation chloride pump NpHR under the control of the excitatory neuron-specific promoter CaMKIIα (AAV-CaMKIIα-ChR2-eYFP or AAV-CaMKIIα-eNpHR3.0-eYFP) was injected into the PL (A/P: +2.7 mm; M/L: ±0.5; D/V: -2.2 mm) using a 10 μl syringe and a thin 31 gauge metal needle with a beveled tip (Hamilton Company). The total injection volume (1 μl) and rate (0.1 μl/min) were controlled with a microinjection pump (UMP3-1, World Precision Instruments). The virus titer was $3 \times 10^{12}$ particles/ml.

### Optrode preparation

For simultaneous extracellular recording and light delivery, we developed a custom-made optrode consisting of a tungsten electrode (1–1.5 MΩ, MicroProbes) attached to an optical fiber (200 μm core diameter, 0.48 NA, Thorlabs). First, a tungsten electrode was attached to a glass capillary tube. An optical fiber was then inserted into the capillary tube and fixed loosely to the electrode using suture thread. Position of the fiber tip was adjusted so that the center-to-center distance between the electrode tip and the fiber tip was ~300 μm.

### In vivo extracellular recording

Each rat transduced in the PL with ChR2 or NpHR was anesthetized with urethane (1.5 g/kg, i.p.), and a small hole was made on the skull above the PL. An optrode was then carefully lowered through the hole until emergence of optically evoked or inhibited signals. Recorded signals were band-pass filtered (0.3–8 kHz), amplified (ExAmp-20K, Kation Scientific), digitized at 20 kHz (NI USB-6009, National Instruments), and stored in personal computer using custom software (LabVIEW, National Instruments). The custom software was also used for controlling a blue (473 nm) or green (532 nm) single diode laser (Shanghai Laser & Optics Century Co.).

### Results

To test the functionality of ChR2-transduced PL, we performed *in vivo* extracellular recordings from PL neurons during delivery of blue light. Indeed, PL neurons showed spiking with perfect fidelity by the blue-light pulse trains at various frequencies (1–20 Hz) in ChR2-transduced PL (n=8, Figure 1a).

> **Photoactivation of a ChR2-transduced PL neuron at various frequencies**
> *157 Data Files*
> (http://dx.doi.org/10.6084/m9.figshare.93270)

Figures 1b and 1c show the results of 10 ms blue-light pulse delivery (<250mW/mm$^2$) at 20 Hz for 5 seconds (100 pulses total). Even when using this longer duration protocol, each light pulse was able to evoke an action potential with high reproducibility. We then recorded 120 minutes (2 hrs) of light-evoked responses in PL neurons (n=2).

> **Repeated photoactivation of a ChR2-transduced PL neuron**
> *133 Data Files*
> (http://dx.doi.org/10.6084/m9.figshare.93271)

The average spontaneous firing rate during the baseline period (5 seconds) during the <u>first</u> 10 minutes of recording was 0.68 +/- 0.10 (Cell 1) and 1.13 +/- 0.19 (Cell 2). Five seconds after delivery of 100 10 ms blue light pulses at 20 Hz the average spontaneous firing rate was 0.56 +/- 0.13 (Cell 1) and 0.067 +/- 0.067 (Cell 2). The average spontaneous firing rate during the baseline period during the <u>last</u> 10 minutes of the 120 minute recording was 0.12 +/- 0.08 (Cell 1) and 1.83 +/- 0.23 (Cell 2) and after blue light delivery the average firing rate was 0.16 +/- 0.074 (Cell 1) and 0.4 +/- 0.15 (Cell 2).

After a baseline period of 5 seconds, trains of 10 ms light pulses (100 pulses total) were repeatedly delivered at 20 Hz with an inter-train interval of 2 min for 120 minutes (Figure 2). Figure 2b shows the high fidelity responsiveness to the light train across the 2 hr recording. Mean spike probabilities (± standard error of the mean) in response to each of the 100 light pulses delivered at 20 Hz for first and last 10 min recording session were as follows: 1.04 ± 0.020 (Cell 1, first) and 0.91 ± 0.034 (Cell 1, last); 2.16 ± 0.055 (Cell 2, first) and 2.19 ± 0.056 (Cell 2, last, Figure 2b).

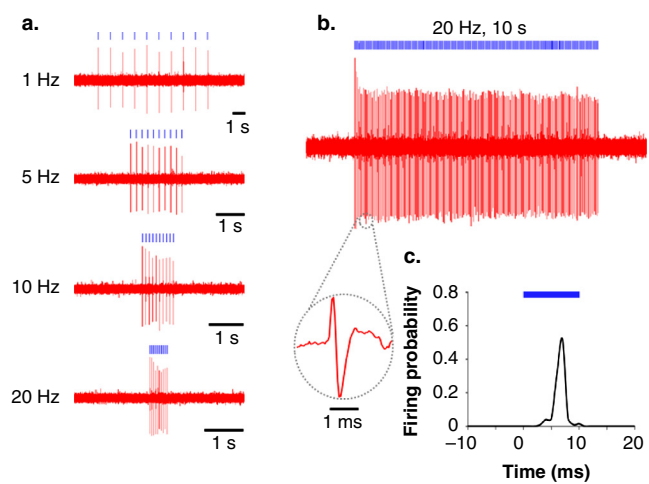

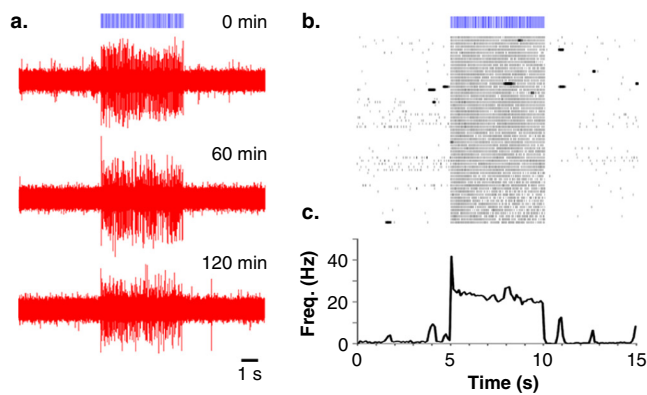

**Figure 1. ChR2 evokes temporally precise unit activity in rat PL cortex *in vivo*. a)** Photoactivation of ChR2-transduced PL neurons at various frequencies (1–20 Hz). **b)** Voltage trace showing blue (473 nm) light-evoked spiking of PL neuron to 20 Hz delivery of 10 ms blue light pulses. Inset: representative light-evoked single-unit response. **c)** Averaged response of this neuron to 10 ms light pulse calculated from all events in above trace (bin width of 1 ms).

**Figure 2. Repeated 20 Hz blue light (10 ms) stimulation of a ChR2-transduced PL neuron. a)** Voltage traces of the light-evoked spiking acquired at the time point of 0, 60 and 120 min after the beginning of recordings. **b)** Raster plot showing all 61 repetitions (2 min interval, 120 min total) of the light-induced activation. **c)** Average firing rate (thick line) with standard error of the mean (SEM, thin line) calculated from raster plot above (bin width of 100 ms).

We then tested *in vivo* photoinhibition of spontaneous activity in NpHR-transduced PL.

**Photoinhibition of NpHR-transduced PL neuron activity**

*61 Data Files*

(http://dx.doi.org/10.6084/m9.figshare.93272)

Figure 3 shows the results of NpHR-induced inhibition of PL neuronal activity using 10 seconds of continuous green light exposure (<250 mW/mm$^2$). In contrast to the ChR2 results, we observed silencing of spontaneous activity in the PL during light delivery (n=7) compared to pre- and post-light periods. Only four (13%, from two cells) out of a total of 30 trials had break through action potentials during the period of 10 second laser illumination (e.g. Figure 3b). To functionally characterize the neuronal activity we constructed autocorrelation histograms (ACH) and calculated the average firing rate during pre- and post-laser period for each neuron (Figure 4). The ACH revealed two types of neurons that were functionally classified into either phasic or non-phasic firing. Phasic neurons showed correlated activity marked by peaks in their ACH (Figure 4a), indicating their periodic network-driven firing pattern, but non-phasic neurons showed no correlated activity (Figure 4b). Two out of three phasic neurons showed a rebound increase in excitability after the light illumination (Figure 4c), but overall there appeared to be no change (Figures 4, c and d). Further recordings are necessary to determine if continuous activation of NpHR with green light leads to some form of plasticity depending on their type of neuronal inputs. Control recordings using blue and green light (light intensities of 250 mW/mm$^2$) did not alter spontaneous firing rates in PL neurons that did not express either ChR2 or NpHR as inferred from the lack of time-locked light responses (data not shown), which suggests no unexpected off-target effects of light alone on neuronal activity.

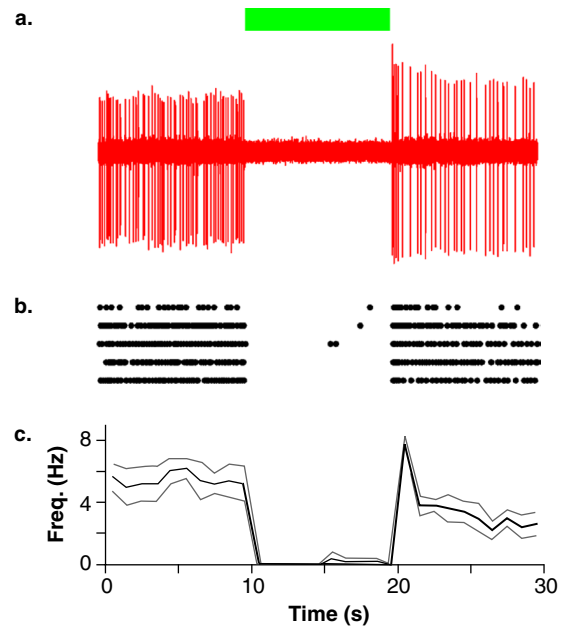

**Figure 3. NpHR rapidly and reversibly silences spontaneous activity of rat PL neurons *in vivo*. a**) Representative trace of spontaneous activity of a NpHR-transduced PL neuron that was inhibited by the continuous green (532 nm) laser delivery. **b**) Raster plot showing five repetitions of the light-induced silencing in this neuron. Each unit activity is plotted as a dot. **c**) Average firing rate (thick line) with SEM (thin line) calculated from these five repetitions (bin width of 1000 ms).

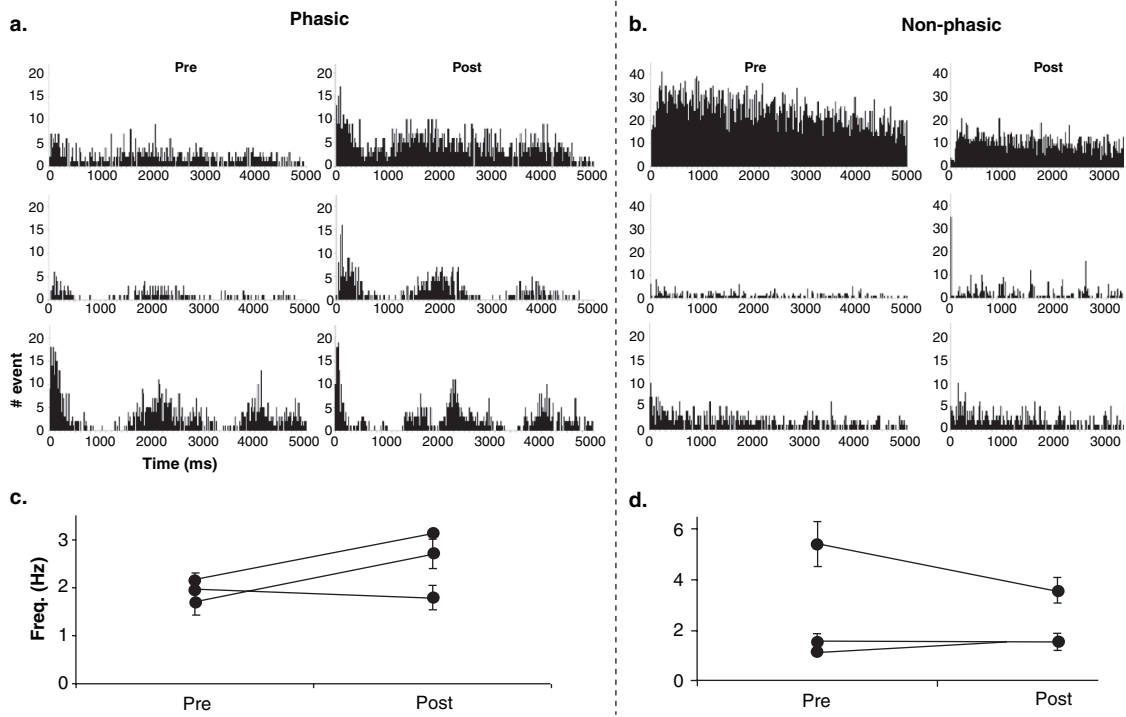

**Figure 4. Comparison of spontaneous activities during pre- and post-laser period for phasic (left) and non-phasic (right) PL neurons. a, b**) Auto-correlation histograms (ACH) for each recorded neuron showing periodic firing pattern in phasic neurons, but not in non-phasic neurons. ACH were created by summing the number of intervals between non-consecutive events as well as intervals between consecutive events (bin width of 20 ms). **c, d**) Average firing rate of each neuron during pre- and post-laser period. Note that the result of one neuron is not shown because of the low number of spikes.

## Discussion

In this study we demonstrated high fidelity *in vivo* recording of individual PL pyramidal neurons transduced with ChR2 or NpHR. Single-unit responses were recorded in response to trains of 10 ms blue light pulses up to 20 Hz with a mode of activation onset time of 7 ms. Light delivered through a custom optrode (<250 mW/mm$^2$) induced stable and robust activation that persisted for the duration of recording (2 hrs in some cases).

To understand how neuronal activity influences behavior it is necessary to not only activate, but also silence neuronal activity with precision. We used 5–10 seconds of green light (<250 mW/mm$^2$) to silence spontaneously active phasic and nonphasic firing NpHR transduced PL pyramidal neurons. Both phasic and nonphasic pyramidal neurons were potently inhibited for the duration of light delivery (up to 10 seconds). In some phasic firing neurons we observed a slight rebound increase in excitability upon termination of light. More recordings are necessary to determine the mechanism for rebound excitability or if it is particular for phasic firing neurons. A transient increase in activity shortly after termination of light-induced hyperpolarization (~100 ms) as seen in Figure 3C would be consistent with activation of a hyperpolarization activated (Ih) inward current, while more persistent increases in activity may indicate indirect effects resulting from activation of the NpHR Cl$^-$ pumps and the resulting change in extracellular Cl$^-$. Recently, Raimondo et al. (2012) reported that *in vitro* synaptically-evoked spike probability significantly increases shortly after termination of photoinhibition in NpHR-expressing hippocampal neurons[8]. Our results are the first to suggest this phenomenon may happen *in vivo* to a subpopulation of phasic firing cells in the PL. Future experiments are necessary to understand the relationship between neuronal activation/silencing on behavior and the possible off target effects of ChR2 or NpHR activation.

The technique of *in vivo* light delivery and simultaneous recording of neuronal responses holds the promise of establishing direct causal relationships between the onset/offset and pattern of activity in specific genetically-determined neuronal populations and corresponding time-locked behavioral events. Our results establish the feasibility of long duration high fidelity activation or continuous silencing of individual neurons for a variety of behavioral experiments.

### Author contributions
S.N., M.V.B, and D.C.C. designed experiments, analyzed data and wrote the paper. S.N., M.V.B., M.B.P., and S.D.D. carried out experiments. M.V.B. and M.B.P. assisted with molecular biology and virus preparation.

### Competing Interests
No relevant competing interests declared.

### Grant information
This work was supported by National Institute on Drug Abuse grant R01-DA24040 (to D.C.C.), NIDA K award K-01DA017750 (to D.C.C.)